\documentclass[journal]{IEEEtran}
\usepackage{mathptmx}
\makeatletter
\newcommand\subparagraph{%
  \@startsection{subparagraph}{5}
  {\parindent}
  {3.25ex \@plus 1ex \@minus .2ex}
  {-1em}
  {\normalfont\normalsize\bfseries}}
\makeatother
\let\subparagraph\relax 
\usepackage[compact]{titlesec}
\titlespacing{\section}{0pt}{2ex}{1ex}
\titlespacing{\subsection}{0pt}{1ex}{0ex}
\titlespacing{\subsubsection}{0pt}{0.5ex}{0ex}
\usepackage{savesym}
\savesymbol{labelindent}
\usepackage[shortlabels]{enumitem}
\setlist[enumerate, 1]{align=left}
\setlist[enumerate]{nosep}
\usepackage{balance}

\usepackage{multirow}
\usepackage{mathtools}
\usepackage{amssymb}
\usepackage{bbold}
\newcommand{\GBmax}[1]{L_{\text{GB},{#1}}^{h,M_{i}}}
\newcommand{\UGB}{\Delta t_{\text{GB}}}
\newcommand{\DtST}{{\Delta}t_{\text{ST}}}
\newcommand{\Lclose}[1]{L^h_{\textit{ST},{#1}}}
\newcommand{\LclosePeriod}{L^h_{\text{ST,GCL}}}
\newcommand{\NGCL}{N^h_{\textit{ST}}}
\newcommand{\pGCL}{p^h_\GCL} 

\newcommand{\test}[1]{{\mathbb 1}_{\set{#1}}}
\newcommand{\set}[1]{\left\{{#1}\right\}}
\newcommand{\st}{~\vrule~} 

\newcommand{\ceil}[1]{\left\lceil#1\right\rceil}

\newcommand{\nnceil}[1]{{\ceil{#1}}^+}
\newcommand{\nnfceil}[2]{\nnceil{\frac{#1}{#2}}}
\newcommand{\isdef}{\ensuremath{\overset{\textit{def}}{=}}}
\newcommand{\nnR}{\mathbb{R}_+}
\usepackage{tikz}

\ifCLASSINFOpdf
\else
\fi

\usepackage{geometry}
\usepackage{url}
\usepackage{color}
\usepackage{xcolor}

\newcommand{\GCL}{\text{GCL}}

\usepackage{graphicx}
\usepackage{subfigure}
\usepackage{float}
\usepackage{stfloats}
\usepackage{amsmath}
\newtheorem{theorem}{Theorem}
\newtheorem{lemma}{Lemma}

\newcommand{\tabincell}[2]{\begin{tabular}{@{}#1@{}}#2\end{tabular}}

\usepackage{authblk}
\begin{document}
%
\title{Latency Analysis of Multiple Classes of AVB \\Traffic in TSN with Standard Credit Behavior \\using Network Calculus}
%
%
%

\author[a]{Luxi Zhao\textsuperscript{*}}
\author[a]{Paul Pop}
\author[b]{Zhong Zheng}
\author[c]{Hugo Daigmorte}
\author[d]{Marc Boyer\thanks{E-mail: luxzha@dtu.dk; paupo@dtu.dk; zhengzhong@buaa.edu.cn; hugo.\\daigmorte@realtimeatwork.com; Marc.Boyer@onera.fr.}}
\affil[a]{Department of Applied Mathematics and Computer Science, Technical University of Denmark, Denmark}
\affil[b]{Department of Electronics and Information Engineering, Beihang University, Beijing, China}
\affil[c]{RealTime-at-Work, Toulouse, France}
\affil[d]{Department on Information Processing and Systems at ONERA, Toulouse, France}
\maketitle

\begin{abstract}
Time-Sensitive Networking (TSN) is a set of amendments that extend Ethernet to support distributed safety-critical and real-time applications in the industrial automation, aerospace and automotive areas. TSN integrates multiple traffic types and supports interactions in several combinations. In this paper we consider the configuration supporting Scheduled Traffic (ST) traffic scheduled based on Gate-Control-Lists (GCLs), Audio-Video-Bridging (AVB) traffic according to IEEE 802.1BA that has bounded latencies, and Best-Effort (BE) traffic, for which no guarantees are provided. The paper extends the timing analysis method to multiple AVB classes and proofs the credit bounds for multiple classes of AVB traffic, respectively under frozen and non-frozen behaviors of credit during guard band (GB). They are prerequisites for non-overflow credits of Credit-Based Shaper (CBS) and preventing starvation of AVB traffic. Moreover, this paper proposes an improved timing analysis method reducing the pessimism for the worst-case end-to-end delays of AVB traffic by considering the limitations from the physical link rate and the output of CBS. Finally, we evaluate the improved analysis method on both synthetic and real-world test cases, showing the significant reduction of pessimism on latency bounds compared to related work, and presenting the correctness validation compared with simulation results. We also compare the AVB latency bounds in the case of frozen and non-frozen credit during GB. Additionally, we evaluate the scalability of our method with variation of the load of ST flows and of the bandwidth reservation for AVB traffic.
\end{abstract}


%
\IEEEpeerreviewmaketitle

\section{Introduction}
%
%
%
%
\IEEEPARstart{E}{thernet} is a well-established network protocol that has excellent bandwidth, scalability, compatibility and cost properties~\cite{IEEE802.3}. However, it was not suitable for real-time and safety critical applications~\cite{Decotignie05}. Distributed safety-critical applications, like those found in the aerospace, automotive and industrial automation domains, require certification evidence for the correct real-time behavior of critical communication. Therefore, several extensions to the Ethernet protocol have been proposed, such as, ARINC 664 Specification Part 7~\cite{ARINC03}, TTEthernet~\cite{SAE11}, and EtherCAT~\cite{Jansen04}. In 2012, the IEEE 802.1 Time-Sensitive Networking (TSN) Task Group~\cite{IEEETSNTG} has been founded to define standard real-time and safety-critical enhancements for Ethernet.

TSN integrates multiple traffic types and supports interactions in several combinations. In this paper we consider a TSN solution supporting Credit-Based Shaper (CBS), previously defined for Audio-Video Bridging (AVB) in 802.1BA~\cite{IEEEBA}, currently in \cite[\S~8.6.8.2]{IEEE802.1Q}, with enhancements for Scheduled Traffic (previously defined in 802.1Qbv~\cite{IEEEQbv}, now in \cite[\S~8.6.8.4]{IEEE802.1Q}) based on Gate-Control Lists (GCLs). We call this architecture TSN/GCL+CBS. We consider three traffic-types with varying criticalities: Scheduled Traffic (ST) which is also known as Time-Triggered (TT) traffic, AVB traffic and Best-Effort (BE) traffic. ST traffic supports hard real-time applications that require very low latency and jitter (e.g., industrial control requires 1--10 milliseconds latency and microseconds jitter). It is transmitted based on schedule tables called GCLs that rely on a global synchronized clock (802.1ASrev~\cite{IEEEASrev}). AVB traffic is intended for applications that require bounded end-to-end latencies, but has less time-critical compared with ST traffic. It uses the CBS to prevent the starvation of lower priority AVB traffic. The TSN Task Group has extended AVB architecture, allowing the definition of multiple CBS~\cite[\S~5.4.1.5]{IEEE802.1Q}, so we consider any number of AVB classes. BE traffic is used for applications that do not require any timing guarantees and is mapped to the remaining queues.

The schedulability of the scheduled ST traffic can be guaranteed during design phase by considering individual scheduling of ST flows and synthesizing the GCLs~\cite{Craciunas16}. For more generic window-based GCLs, Zhao et al.~\cite{LuxiAccess18} have proposed a Network Calculus (NC)-based method to analyze the latency bounds for flexible window-based ST traffic. Researchers have extended the analysis to the preemption mode using Real-Time Calculus~\cite{ZhangAccess19}. Moreover, AVB in TSN can be used for time-critical applications that require bounded latency even though it has lower time-critical than ST traffic. Hence it is very important to have a safe and guaranteed WCD analysis for AVB that takes into account the ST flows. Although latency analysis methods have been successfully applied to AVB traffic in AVB networks~\cite{Azua14, Philip14, NC-AVB-3classes}, they do not take account for the interference of ST traffic on the AVB traffic and only consider two or three AVB classes. 
Recently, Cao et al.~\cite{FA-EI-AVB-Mclasses} have proposed an eligible interval based formal analysis on the calculation of worst-case performance bounds for multiple classes of AVB traffic. However, they do not consider the effect of ST traffic on AVB traffic, and focus only on the ``relative delay'' on a single hop. Network Calculus-based analyses have been proposed~\cite{Zhao17} to compute the WCDs of Rate-Constrained (RC) traffic with the consideration of the ST (also named Time-Triggered, or TT) frames in TTEthernet, but the techniques are not applicable for TSN: RC traffic has no CBS, the integration modes are different in TSN and TTEthernet and TSN schedules ST traffic differently from TTEthernet: individual ST frames are considered in TTEthernet, whereas TSN schedules ST windows which may include several ST frames.


The timing analysis of non-ST traffic types in TSN has been addressed in \cite{Thiele15}, considering closed-gate blocking, strict priorities and a ``Peristaltic Traffic Shaper''. However, their analysis is not applicable to AVB, which uses CBS. Researchers~\cite{Mohammadpour18},\cite{Zhang19} have proposed the latency bounds for AVB traffic affected by control-data-traffic (CDT) in TSN, but assuming CDT as leaky bucket (LB) or length rate quotient (LRQ), which does not fit for the ST traffic. The AVB Latency Math equation has been extended to consider the ST traffic in TSN~\cite{Laursen16}. However, it does not consider the actual situations for AVB flows in the network, but just assumes that maximum allowable bandwidth is occupied by the corresponding AVB traffic class, thus causing overly pessimistic, i.e., leading to overly large WCDs. In addition, \cite{Laursen16} can only be used to determine the WCDs of AVB Class A traffic. Researchers have extended the Eligible Interval Analysis (EIA) to calculate the delay of two classes of AVB traffic in TSN~\cite{Maxim17}. However, it does not consider relative offsets between ST windows, but just assumes the ST windows always arranging back-to-back. The initial idea for the two classes of AVB analysis under the influence of ST traffic in TSN network based on the Network Calculus has been given by \cite{ZhaoRTAS17}, and then credit bounds for CBS are improved in~\cite{Mohammadpour19}. However, all these analyses above are based on a credit behavior deviating from the standard 802.1Q-2018~\cite{IEEE802.1Q}, i.e., assuming credit frozen during guard band before gate closing. Moreover, the latest standard 802.1Q supports any number of AVB classes, which are not supported in previous analysis work. Thus, in this paper, we are interested to propose a formal performance analysis method for arbitrary number of AVB classes in TSN networks, under both behaviors of frozen and non-frozen credit during guard band (GB, cf. Sect.~\ref{sec:TSN Protocol} for details).

The main contributions of our paper are as follows:
\begin{enumerate}[leftmargin=4pt]
\item[$\bullet$] To the best of our knowledge, our analysis is the first one to prove the credit bounds of CBS in the cases of frozen and non-frozen credit during guard band. The analysis is able to handle with arbitrary number of AVB classes under the influence of ST traffic for the whole TSN network.
\item[$\bullet$] We reduce the pessimism of the analysis and this provides tighter latency bounds for AVB, by introducing the limitations from the physical link rate and the output of CBS, which are denoted as the link and CBS shaping curves with the consideration of ST traffic.
\item[$\bullet$] We evaluate the proposed approach on both synthetic and real-world test cases, comparing AVB latencies in cases of frozen and non-frozen credit during guard bands, and comparing with related work and simulation results to show the significant reduction of pessimism on latency bounds, and to validate the correctness and the scalability of our implementation.
\end{enumerate}

The paper is organized as follows. Sect.~\ref{sec:System Model} presents the system model. Sect.~\ref{sec:TSN Protocol} introduces TSN. Sect.~\ref{sec:Network Calculus Background} briefly introduces the Network Calculus concepts needed for the analysis. Sect.~\ref{sec:Worst-case Analysis for AVB Traffic} gives the proof of credit bounds for multiple classes of AVB traffic respectively under frozen and non-frozen behaviors of credit during GB and presents tighter AVB latency bounds by introducing shaping curves. Sect.~\ref{sec:Experimental Results} evaluates the proposed analysis and Sect.~\ref{sec:Conclusion and Future Work} concludes the paper.

\section{System Model}
\label{sec:System Model}

A TSN network is composed of a set of end systems (ES) and switches (SW) also called nodes, connected via physical links. In this paper, we assume, without loss of generality, that all physical links have the same rate $C$. The links are full duplex, allowing thus communication in both directions, and the networks can be multi-hop. The output port of a SW is connected to one ES or an input port of another SW. 

The messages are sent from ESes via flows, which have a single source and may have multiple destinations. Each source ES is able to send multiple flows to the network. As mentioned, our TSN solution supports three traffic types: ST~\cite[\S~8.6.8.4]{IEEE802.1Q}, AVB~\cite[\S~8.6.8.2]{IEEE802.1Q} and BE. We assume that the traffic type for each application has been decided by the designer.
The subscripts $M_i$ ($i\in [1, 6]$) for AVB denote respectively the different AVB traffic classes. For ST traffic, we know the GCLs in each output port $h$ of nodes, i.e., the opening and closing time of ST traffic (named as ST window) and GCL period ($p_{\GCL}^h$). For an AVB flow $\tau_{M_i[k]}\in\tau_{M_i}$, we know its frame size $l_{M_i[k]}$, the minimum frame interval $p_{M_i[k]}$ in the source ES and the traffic class $M_i$ it belongs to. The AVB Class $M_i$ has higher priority than the AVB Class $M_{i+1}$. The flows assigned the same AVB traffic class $M_i$ are served in FIFO order. Moreover, we know the maximum frame size $l_{BE}^{max}$ of BE traffic.

\section{TSN/GCL+CBS Output Port}
\label{sec:TSN Protocol}

\begin{figure}[!t]
 \centering
 \includegraphics[width=0.294\textwidth]{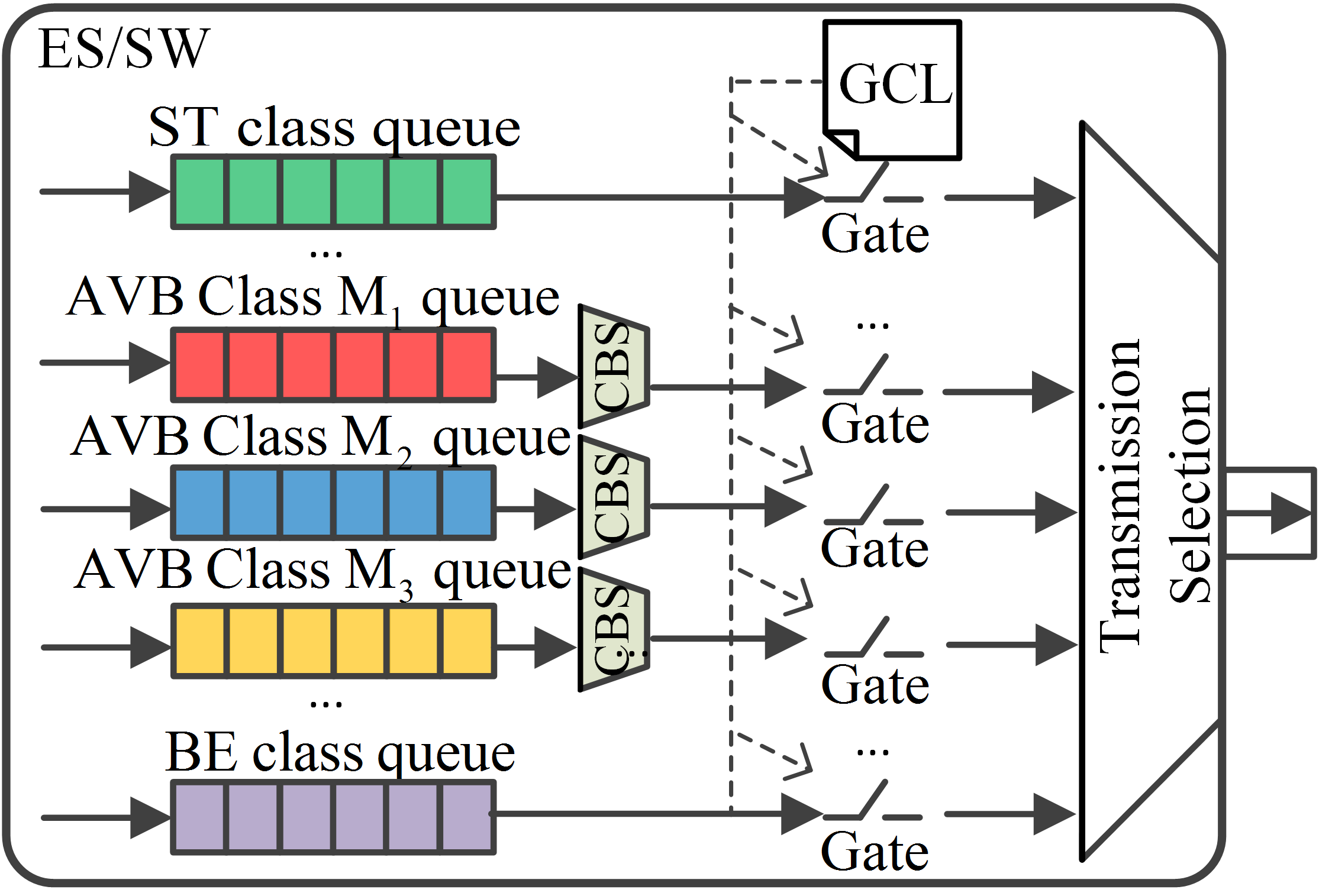}
 \caption{\label{fig:TAS} TSN/GCL+CBS architecture for an output port in an ES/SW}
\end{figure}

The TSN Task group has defined a set of schedulers, which can interact in several combinations. In this paper, we assume a specific configuration, called TSN/GCL+CBS architecture to provide different real-time guarantees and will present how ST and AVB flows are transmitted in this case, as shown in Fig.~\ref{fig:TAS}. Each one has eight queues for storing frames that wait to be forwarded on the corresponding link, one or more for ST queues, two or more ($n_{CBS}^h$) for AVB queues (respectively for Class $M_i$) and the remaining queues are used for BE. Every queue has a gate with two states, open and closed. Frames waiting in the queue are eligible to be forwarded only if the associated gate is open.

The gates for each queue are controlled by GCLs, which are created offline and contain the times when the associated gates are open and closed~\cite{Craciunas16}. In this study, we assume that when associated gate for ST traffic is open, the remaining gates for other traffic (AVB and BE) are closed, and vice versa (aka. exclusive gating). Therefore, AVB traffic is prevented from transmitting in the time windows reserved for ST frames.

\begin{figure}[!t]
 \centering
 \includegraphics[width=0.33\textwidth]{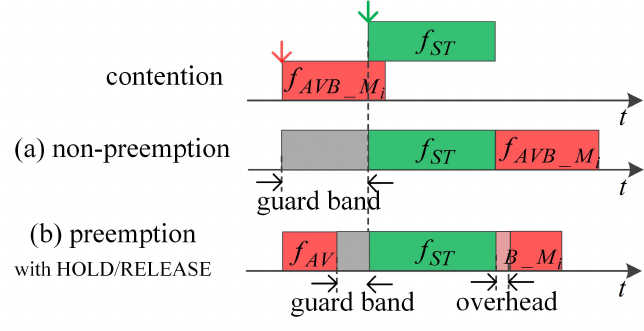}
 \caption{\label{fig:IntegrationModes} Non-preemption integration modes}
\end{figure}

In addition, two integration modes are introduced to solve the issue when an AVB frame is already in transmission at the beginning of time window for ST, i.e., non-preemption~\cite[Annex Q]{IEEE802.1Q} and preemption~\cite[Annex S]{IEEE802.1Q} modes. In this paper, we focus on the discussion of the \textit{non-preemption} integration mode (see Fig.~\ref{fig:IntegrationModes}) and the analysis model is also suitable for the preemption integration mode combined with HOLD/RELEASE. The non-preemption mode uses a ``guard band'' before each ST window to prevent the AVB frame from initiating transmission if there is insufficient time available to transmit that entire frame before the open event of ST gate. The preemption mode combined with HOLD/RELEASE also uses a guard band, but with a rather smaller size~\cite[Annex S.4]{IEEE802.1Q}.
The ``guard band'' will lead to wasted bandwidth due to the guard band, but it ensures no delay for ST traffic.

\begin{figure}[!t]
	\centering
	\includegraphics[width=0.401\textwidth]{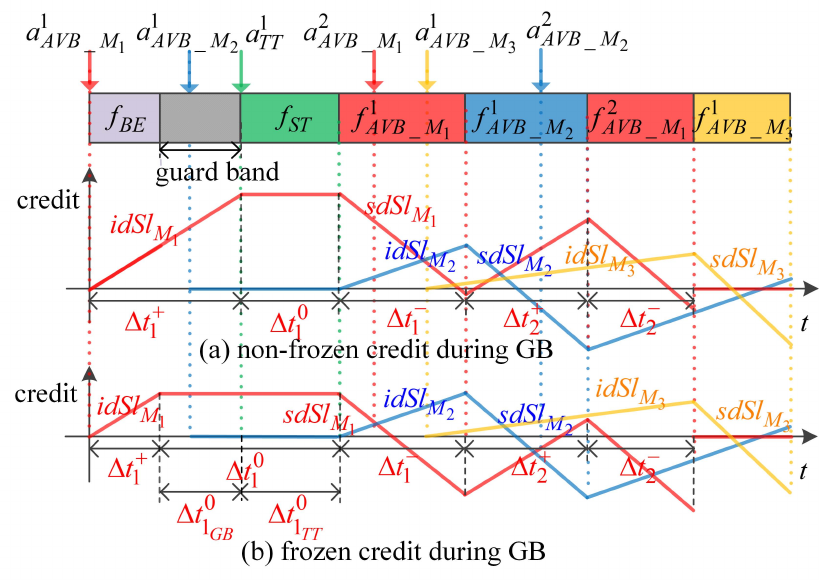}
	\caption{\label{fig:CBS} CBS example with non-preemption mode}
\end{figure}

An enqueued AVB frame is transmitted only if the associated gate is open and the Credit-Based Shaper (CBS)~\cite[\S~8.6.8.2]{IEEE802.1Q} permits it. Each AVB queue has a credit value, initialized to zero. When the AVB gate is closed, the associated credit is ``frozen''. When the associated AVB gate is open, the variation of credit is the same as the one in AVB networks~\cite{IEEEBA}, i.e., decreasing with a send slope ($sdSl_{M_i}$) during the transmission of an AVB frame and increasing with an idle slope ($idSl_{M_i}$) when AVB frames are waiting to be transmitted due to other higher priority AVB frames or the negative credit. This is illustrated for example in Fig.~\ref{fig:CBS}, where we have three AVB classes $M_i$ ($i=1, 2, 3$), and show the variation of credit for respective AVB class. Particularly, during guard bands, we assume two behaviors of the credit for AVB traffic. One is ``non-frozen'', obeying the behavior of credit in the standard. The other is ``frozen'', of which the behavior deviating from the standard, but is a model assumption for most existing work. Daigmorte et al.~\cite{Daigmorte19} have shown that the credit evolution rule with guard band in the standard is unfair, which is an argument for an update of the standard to consider the credit evolution rule to be frozen.

\section{Network Calculus Background}
\label{sec:Network Calculus Background}
Network Calculus~\cite{LeBoudec01} is a mature theory proposed for deterministic performance analysis. It is used to construct arrival and service curve models for the investigated flows and network nodes. Network Calculus functions mainly belong to non-decreasing functions and null before 0: $\mathcal{F}_\uparrow=\{f:\mathbb{R}_+\rightarrow\mathbb{R}|x_1<x_2\Rightarrow f(x_1)<f(x_2), x<0\Rightarrow f(x)=0\}$. Two basic operators on $\mathcal{F}_\uparrow$ are the convolution $\otimes$,
\begin{equation}\label{g:conv}
(f\otimes g)(t)=\inf_{0\leq s\leq t}\{f(t-s)+g(s)\},
\end{equation}
and deconvolution $\oslash$,
\begin{equation}\label{g:deconv}
(f\oslash g)(t)=\sup_{s\geq0}\{f(t+s)-g(s)\},
\end{equation}
where $\inf$ means infimum and $\sup$ means supremum.

An arrival curve $\alpha(t)$ is a model constraining the arrival process $R(t)$ of a flow, in which $R(t)$ represents the input cumulative function counting the total data bits of the flow that has arrived in the network node up to time $t$. We say that $R(t)$ is constrained by $\alpha(t)$ iff for all $s\leq t$,
\begin{equation}\label{g:arr}
R(t)-R(s)\leq\alpha(t-s).
\end{equation}
\noindent

A service curve $\beta(t)$ models the processing capability of the available resource. Assume that $R^*(t)$ is the departure process, which is the output cumulative function that counts the total data bits of the flow departure from the network node up to time $t$. There are several definitions for service curve. We say that the network node offers the min-plus minimum service curve $\beta(t)$ for the flow iff
\begin{equation}\label{g:2}
R^{*}(t)\geq \inf_{0\leq s\leq t}\left\{R(s)+\beta(t-s)\right\}=(R\otimes \beta)(t),
\end{equation}

\noindent
and offers the strict service curve $\beta(t)$ iff
\begin{equation}\label{g:strSC}
R^{*}(t+\Delta t)-R^{*}(t)\geq \beta(\Delta t),
\end{equation}

\noindent
during any backlog period $(t,t+\Delta t]$. In addition, in order to evaluate service curves we will use the non-decreasing non-negative closure defined by
\begin{equation*}
[f(t)]_\uparrow^{+}=\max \limits_{0 \leq s \leq t} \{f(s),0\}.
\end{equation*}

A shaping curve $\sigma(t)$ characterizes the maximum number of bits that are served during a period of time $\Delta t$, which means that the departure process $R^*(t)$ from the server always respects the shaping curve. A server offers a shaping curve $\sigma(t)$ iff $\sigma(t)$ could be an arrival curve for all output cumulative function $R^*(t)$ (Eq.~\ref{g:arr}).
Note that the shaping curve $\sigma(t)$ from the output can also be said as an arrival curve for the subsequent nodes.

If a flow $R(t)$ of arrival curve $\alpha(t)$ crosses a server with the service curve $\beta(t)$, then the output flow $R^*(t)$ can be bounded by the arrival curve $\alpha'(t)$,
\begin{equation}\label{g:3}
\alpha'(t)=\alpha\oslash\beta(t)=\sup_{s\geq 0}\left\{\alpha(t+s)-\beta(s)\right\},
\end{equation}
\noindent
It can also be taken as an arrival curve of input flow for the next node.

Let us assume that the flow constrained by the arrival curve $\alpha(t)$ traverses the network node offering the service curve $\beta(t)$. Then, the latency experienced by the flow in the network node is bounded by the maximum horizontal deviation between the graphs of two curves $\alpha(t)$ and $\beta(t)$,
\begin{equation}\label{g:4}
h(\alpha,\beta)=\sup_{s\geq0}\left\{\inf\left\{\tau\geq0\mid\alpha(s)\leq\beta(s+\tau)\right\}\right\}.
\end{equation}
The network end-to-end delay of a frame along its route is defined as the time duration between the instant it is transmitted by the source ES on the link and the instant it is fully received by the destination ES. The worst-case end-to-end delay of the flow can be bounded by the sum of latency bounds in each network nodes along its route.

\section{Worst-case Analysis for AVB Traffic}
\label{sec:Worst-case Analysis for AVB Traffic}
\subsection{Service Curve for AVB Traffic}
\label{subsec:Service Curve for AVB traffic}
In this section, we will extend the service curve in \cite{ZhaoRTAS17} to multiple AVB Class $M_i$ ($i \in [1,n_{CBS}^h]$) with respectively (i) non-frozen and (ii) frozen credit during guard band.
The service curve for AVB traffic depends on the working mechanism of CBS. As can be seen from Fig.~\ref{fig:CBS}, no matter the case (i) or (ii) for each AVB traffic Class $M_i$, any time interval $\Delta t$ can be decomposed into
\begin{equation}\label{g:9}
\Delta t=\Delta t^{+}+\Delta t^{-}+\Delta t^0,
\end{equation}
\noindent
where $\Delta t^+=\sum_i\Delta t_i^+$ (resp. $\Delta t^-=\sum_j\Delta t_j^-$) represents the accumulated length of all period where the credit increases (resp. decreases), and $\Delta t^0=\sum_k\Delta t_k^0$ is the frozen time of credit. 
The service could only be supplied for AVB traffic during the descent time $\Delta t^-$ of credit. Then, the service curve for multiple classes of AVB traffic is given by the following theorem.
\begin{theorem}\label{nprSerAVB}
	The min-plus minimum service curve for AVB Class $M_i$ ($i\in [1,n_{CBS}^h]$) under non-preemption mode in an output port $h$ is given by
	\begin{equation}\label{g:SC_AVB}
	\begin{split}
	\beta_{M_i(I)}^{h}(t)=
	idSl_{M_i}\bigg[ t -\frac{\alpha_{H(I)}^h(t)}{C} -
	\frac{c_{M_i(I)}^{max}}{idSl_{M_i}} \bigg]_\uparrow^+,
	\end{split}
	\end{equation}

	\noindent
	where $I\in\{NF,F\}$ represents the behavior of credit during GB, and $\alpha_{H(I)}^h(t)$ is the arrival curve with regard to credit frozen. In case of $I=NF$, i.e., (i) non-frozen credit during GB, $\alpha_{H(I)}^h(t)=\alpha_{ST}^h(t)$ is only related to higher priority ST traffic, while in case of $I=F$, i.e., (ii) frozen credit during GB, $\alpha_{H(I)}^h(t)=\alpha_{GB+ST}^h(t)$ should additionally take guard band slots into account. $\alpha_{ST}^h(t)$ and $\alpha_{GB+ST}^h(t)$ are respectively given by Lemma~\ref{nprAggArrST} and Lemma~\ref{nprAggArrSTGB} in Sect.~\ref{subsec:AC_ST/GB}. $c_{M_i(I)}^{max}$ is the upper bound of credit of Class $M_i$ under cases of (i) non-frozen or (ii) frozen credit during GB, and given by Eqs. (\ref{g:upper_boundNF}) or (\ref{g:upper_boundF}). The proof of credit bound has to be extended to an arbitrary number of AVB classes and to consider different behaviors of credit during guard band, which is one of the challenges in this paper and will be discussed in Sect.~$\ref{subsec:Non-overflow condition}$. The proof Theorem~\ref{nprSerAVB} is similar as the one given in \cite{ZhaoRTAS17}, hence it will not be further discussed in this paper.
\end{theorem}

\subsection{Bounding the Credit for AVB Traffic}
\label{subsec:Non-overflow condition}
In this section, we bound the credit for AVB traffic. Let us recall from Sect.~\ref{sec:TSN Protocol} how AVB is transmitted.

\begin{theorem}\label{credit_lower_bound}
(Lower bound of Class $M_i$) Let $l_{M_i}^{max}$ be the maximal frame size of any flow crossing the AVB queue $Q_{M_i}$. Then, the credit $c_{M_i}(t)$ of Class $M_i$ is lower bounded by \cite{Azua14,ZhaoRTAS17},
\begin{equation}\label{g:lower_bound}
c_{M_i}(t)\geq \frac{l_{M_i}^{max}}{C} sdSl_{M_i}= c_{M_i}^{min}.
\end{equation}
\end{theorem}




Before we start to prove the upper bound of credit, let us define some notations. Let $Q_{\textit{AVB}}^{\leq i}=\bigcup_{1\leq j\leq i}\{Q_{M_j}\}$ denotes the same or higher priority AVB queues and $Q_{AVB}^{> i}=\bigcup_{i< j\leq n_{CBS}^h}\{Q_{M_j}\}$ denotes the lower priority AVB queues. Then for $\forall s, t\in\mathbb{R}^+, s\leq t$, the interval $[s,t]$ can be partitioned into several kinds of intervals: let $\Delta t_{M_i}(s,t)$ and $\Delta t_{AVB}^{<i}(s,t)$ respectively denote the duration of emissions of frames from the queue $Q_{M_i}$ and of high priority frames from the queue $Q_{AVB}^{<i}$; $\Delta t_{ST}(s,t)$ is the time slot reserved for ST traffic, \emph{i.e.}, the duration where the gates of AVB queues are closed; $\Delta t_{LP}(s,t)$ is the duration where a frame in $Q_{M_j}\in Q_{AVB}^{\leq i}$ waits due to a lower priority frame in queue $Q_{AVB}^{>i}$ or $Q_{BE}$ that is being transmitted cannot be preempted; $\Delta t_{GB}(s,t)$ denotes the guard band duration where for all queues $Q_{M_j}\in Q_{AVB}^{\leq i}$ with $c_{M_j}>0$ no frame can be sent, as there is no sufficient time available to transmit the entire frame before the next closing event; and $\Delta t_{Other}(s,t)$ is the interval except all the definitions above. Then we have $t-s= \Delta t_{M_i}(s,t) + \Delta t_{AVB}^{<i}(s,t) + \Delta t_{ST}(s,t) + \Delta t_{LP}(s,t) + \Delta t_{GB}(s,t) +\Delta t_{Other}(s,t)$.
\begin{theorem}\label{credit_upper_bound}
(Higher bound of Class $M_i$) Let $l_{BE}^{max}$ be the maximal frame size of a BE flow. The credit $c_{M_i}(t)$ of Class $M_i$ is upper bounded by,
\begin{enumerate}[(i)]
	\item non-frozen credit during GB
	\begin{equation}\label{g:upper_boundNF}
	c_{M_i}(t)\leq idSl_{M_i}\cdot\frac{\sum_{j=1}^{i-1}c_{M_j}^{min}-l_{>i}^{max}-\sigma_{GB}^{M_i}}{\rho_{GB}^{M_i}+\sum_{j=1}^{i-1}idSl_{M_j}-C}=c_{M_i(NF)}^{max};
	\end{equation}
	\item frozen credit during GB
	\begin{equation}\label{g:upper_boundF}
	c_{M_i}(t)\leq idSl_{M_i}\cdot\frac{\sum_{j=1}^{i-1}c_{M_j}^{min}-l_{>i}^{max}}{\sum_{j=1}^{i-1}idSl_{M_j}-C}=c_{M_i(F)}^{max}.
	\end{equation}
\end{enumerate}
where $l_{>i}^{max}=max_{j\in[i+1,n_{CBS}^h]}\{l_{M_j}^{max},l_{BE}^{max}\}$, $c_{M_j}^{min}$ is the lower bound of credit of Class $M_j$ from Theorem~\ref{credit_lower_bound}, and $\sigma_{GB}^{M_i}$ and $\rho_{GB}^{M_i}$ are the parameters of upper envelope related to guard band duration and satisfy for $\forall s,t\in\mathbb{R}^+, s\leq t$,
\begin{equation}\label{g:deltaTGB}
C\cdot\Delta t_{GB}(s,t)\leq \sigma_{GB}^{M_i}+\rho_{GB}^{M_i}\cdot(t-s-\Delta t_{ST}(s,t)).
\end{equation}
which will be discussed in the end of Sect.\ref{subsec:AC_ST/GB}.
\end{theorem}

\emph{Proof}: Considering the evolution of the credit, the maximal value of credit will only happen at the instant satisfying the following definition. Let $t\in \mathbb{R}^+$ be a time point when AVB gates are in the open state, and $c_{M_i}(t)>0$.
Then let us define $s=\sup{\left\{u\leq t~\vrule~\forall Q_{M_j}\in Q_{AVB}^{\leq i}, c_{M_j}(t)\leq 0\right\}}$.
It implies that $\forall u\in(s,t]$, $\exists Q_{M_j}\in Q_{AVB}^{\leq i}$, $c_{M_j}(u)>0$, \emph{i.e.}, there always exists at least one queue in $Q_{AVB}^{\leq i}$ with some frame to send. Otherwise, we can always find another $s<s'\leq t$ that satisfies $\forall Q_{M_j}\in Q_{AVB}^{\leq i}, c_{M_j}(s')\leq 0$. As it always exists one queue in $Q_{AVB}^{\leq i}$ with some frame to send, either it succeeds (there is a frame of $M_i$ emission, or a high-priority frame in $Q_{M_j}\in Q_{AVB}^{>i}$ emission) or is blocked (non-preemption or guard band or ST traffic transmission). Thus $\Delta t_{Other}(s,t)=0$ here. In the following, as $s, t$ are fixed, the $\Delta t_X(s,t)$ will be simplified to $\Delta t_X$.


(i) \textit{For the case of non-frozen credit during GB}

Consider first the evolution of the credit value of $M_i$ between $s$ and $t$. 
The credit $c_{M_i}(t)$ increases at speed $idSl_{M_i}$ when the frame in the queue $Q_{M_i}$ is waiting (during $\Delta t_{AVB}^{<i}+\Delta t_{LP}+\Delta t_{GB}$), decreases at speed $sdSl_{M_i}=idSl_{M_i}-C$ when the frame in the queue $Q_{M_i}$ is transmitting (during $\Delta t_{M_i}$), is not modified when the gate of $Q_{M_i}$ is closed during ST traffic transmission, and may be reduced from some positive value $P$ to 0 due to resets. Thus the variation of $c_{M_i}(t)$ during $(s,t]$ is,
\begin{equation}\label{g:deltaMi(i)1}
\begin{split}
&c_{M_i}(t)-c_{M_i}(s)\\
&=\Delta t_{M_i}\cdot sdSl_{M_i}+\left(\Delta t_{AVB}^{<i}+\Delta t_{LP}+\Delta t_{GB}\right)\cdot idSl_{M_i}-P.
\end{split}
\end{equation}
Since $\Delta t_{AVB}^{<i}+\Delta t_{LP}+\Delta t_{GB}=s-t-\Delta t_{M_i}-\Delta t_{ST}$ and $P\geq0$, Eq.~(\ref{g:deltaMi(i)1}) is modified into,
\begin{equation}\label{g:deltaMi(i)}
\begin{split}
c_{M_i}(t)-c_{M_i}(s)\leq -\Delta t_{M_i}\cdot C+\left(t-s-\Delta t_{ST}\right)\cdot idSl_{M_i}.
\end{split}
\end{equation}

\hspace{0.2em} Let $c_{<i}(t)=\sum_{j=1}^{i-1}c_{M_j}(t)$ denote the sum of credits of AVB traffic with the priority higher than $M_i$. Consider three cases: first, at any instant between $s$ and $t$ either a frame of $M_i$ uses the link, or a low priority frame blocks the link or during GB, the credit of each non-empty queue $Q_{M_j}\in Q_{AVB}^{<i}$ increases with speed $idSl_{M_j}$. Then the sum of credits $c_{<i}(t)$ increases at most at speed $\sum_{j=1}^{i-1}idSl_{M_j}$. 
Or a frame from class with higher priority than $M_i$ is being sent. In this case $c_{<i}(t)$ decrease at least at speed $\sum_{j=1}^{i-1}idSl_{M_j}-C$ (all the classes from $Q_{AVB}^{<i}$ gain credit expect one which loses credit). The last case is like for $M_i$, $c_{<i}(t)$ may decrease due to a set of resets. Then it is possible to upper bound the variation of $c_{<i}(t)$ between $s$ and $t$,
\begin{equation}\label{g:c<i1(i)}
\begin{split}
&c_{<i}(t)-c_{<i}(s)\leq\left(\Delta t_{M_i}+\Delta t_{LP}+\Delta t_{GB}\right)\cdot \sum_{j=1}^{i-1}idSl_{M_j}\\
&+\left(t-s-\Delta t_{ST}-\Delta t_{M_i}-\Delta t_{LP}-\Delta t_{GB}\right)\cdot \bigg(\sum_{j=1}^{i-1}idSl_{M_j}-C\bigg)\\
&=\left(\Delta t_{M_i}+\Delta t_{LP}+\Delta t_{GB}\right)\cdot C+(t-s-\Delta t_{ST})\cdot \bigg(\sum_{j=1}^{i-1}idSl_{M_j}-C\bigg).
\end{split}
\end{equation}
By definition of $s$, it exists continuously some queue $Q_{M_j}\in Q_{AVB}^{\leq i}$ trying to send a frame. Hence there is at most one low priority frame that can get access to the link before $s$ and go on transmission due to non-preemption. Thus
\begin{equation*}
\Delta t_{LP}\cdot C\leq\max_{j\in[i+1,n_{CBS}^h]}\{l_{M_j}^{max},l_{BE}^{max}\}=l_{>i}^{max}.
\end{equation*}
Moreover, we assume that there exists $\sigma_{GB}^{M_i}$ and $\rho_{GB}^{M_i}$ which will be discussed in Sect.~\ref{subsec:AC_ST/GB} such that $\forall s, t\in\mathbb{R}^+, s\leq t$,
\begin{equation*}
\Delta t_{GB}\cdot C\leq \sigma_{GB}^{M_i}+\rho_{GB}^{M_i}\cdot(t-s-\Delta t_{ST}).
\end{equation*}
Thus Eq.~(\ref{g:c<i1(i)}) is modified to obtain,
\begin{equation}\label{g:c<i2}
\begin{split}
c_{<i}(t)-c_{<i}(s)\leq&\Delta t_{M_i}\cdot C+l_{>i}^{max}+\sigma_{GB}^{M_i}\\
+&(t-s-\Delta t_{ST})\cdot \bigg(\rho_{GB}^{M_i}+\sum_{j=1}^{i-1}idSl_{M_j}-C\bigg).
\end{split}
\end{equation}
Considering the system is assumed to be not overloaded, we have $\rho_{GB}^{M_i}+\sum_{j=1}^{i-1}idSl_{M_j}\leq C$. Thus, from Eq.~(\ref{g:c<i2})
\begin{equation}\label{g:t-s-deltaTST}
t-s-\Delta t_{ST} \leq \frac{c_{<i}(t)-c_{<i}(s)-\Delta t_{M_i}\cdot C-l_{>i}^{max}-\sigma_{GB}^{M_i}}{\rho_{GB}^{M_i}+\sum_{j=1}^{i-1}idSl_{M_j}-C}.
\end{equation}


Then from (\ref{g:t-s-deltaTST}), (\ref{g:deltaMi(i)}) is modified to obtain,
\begin{equation*}
\begin{split}
c_{M_i}(t)-c_{M_i}(s)\leq 
idSl_{M_i}\frac{c_{<i}(t)-c_{<i}(s)-l_{>i}^{max}-\sigma_{GB}^{M_i}}{\rho_{GB}^{M_i}+\sum_{j=1}^{i-1}idSl_{M_j}-C}.
\end{split}
\end{equation*}
By definition of s, we have $c_{<i}(s)\leq 0$ and $c_{M_i}(s)\leq 0$ as well. Moreover, if $c_{M_j}^{min}$ is the lower bound of credit of Class $M_j$, then $c_{<i}(t)\geq\sum_{j=1}^{i-1}c_{M_j}^{min}$ so to conclude,
\begin{equation*}
c_{M_i}(t)\leq idSl_{M_i}\cdot\frac{\sum_{j=1}^{i-1}c_{M_j}^{min}-l_{>i}^{max}-\sigma_{GB}^{M_i}}{\rho_{GB}^{M_i}+\sum_{j=1}^{i-1}idSl_{M_j}-C}.
\end{equation*}

(ii) \textit{For the case of frozen credit during GB}

The proof is similar to the above process, except the difference of the evolution of credit of $M_i$ during GB. The credit $c_{M_i}(t)$ increases at speed $idSl_{M_i}$ during $\Delta t_{AVB}^{<i}+\Delta t_{LP}$, decreases at speed $sdSl_{M_i}=idSl_{M_i}-C$ during $\Delta t_{M_i}$, and is not modified during guard band $\Delta t_{GB}$ and ST traffic transmission $\Delta t_{ST}$. Then the variation of $c_{M_i}(t)$ during $(s,t]$ is upper bounded by,
\begin{equation}\label{g:deltaMi(ii)}
\begin{split}
c_{M_i}(t)-c_{M_i}(s)\leq
-\Delta t_{M_i}\cdot C+(t-s-\Delta t_{GB}-\Delta t_{ST})\cdot idSl_{M_i}.
\end{split}
\end{equation}

Additionally, for the sum of credits $c_{<i}(t)$ of AVB traffic with the priority higher than $M_i$, it increases at most as speed $\sum_{j=1}^{i-1}idSl_{M_j}$ when the frame of $M_i$ uses the link (during $\Delta t_{M_i}$) or a low priority frame blocks the link (during $\Delta t_{LP}$),
\begin{equation}\label{g:c<i1(ii)}
\begin{split}
&c_{<i}(t)-c_{<i}(s)\leq\left(\Delta t_{M_i}+\Delta t_{LP}\right)\cdot \sum_{j=1}^{i-1}idSl_{M_j}\\
&+(t-s-\Delta t_{GB}-\Delta t_{ST}-\Delta t_{M_i}-\Delta t_{LP})\cdot \bigg(\sum_{j=1}^{i-1}idSl_{M_j}-C\bigg)\\
&\leq\Delta t_{M_i}\cdot C+l_{>i}^{max}+\left(t-s-\Delta t_{GB}-\Delta t_{ST}\right)\cdot \bigg(\sum_{j=1}^{i-1}idSl_{M_j}-C\bigg).
\end{split}
\end{equation}
Similarly, considering the system is assumed to be not overloaded, we have $\sum_{j=1}^{i-1}idSl_{M_j}\leq C$. Thus, from Eq.~(\ref{g:c<i1(ii)})
\begin{equation}\label{g:t-s-deltaTGB-deltaTST}
t-s-\Delta t_{GB}-\Delta t_{ST}\leq\frac{c_{<i}(t)-c_{<i}(x)-\Delta t_{M_i}\cdot C-l_{>i}^{max}}{\sum_{j=1}^{i-1}idSl_{M_j}-C}.
\end{equation}

Then from (\ref{g:t-s-deltaTGB-deltaTST}), (\ref{g:deltaMi(ii)}) is modified to obtain,
\begin{equation*}
\begin{split}
c_{M_i}(t)&-c_{M_i}(s)\leq 
idSl_{M_i}\frac{c_{<i}(t)-c_{<i}(x)-l_{>i}^{max}}{\sum_{j=1}^{i-1}idSl_{M_j}-C}.
\end{split}
\end{equation*}
Thus, the upper bound of $c_{M_i}(t)$ is to conclude,
\begin{equation*}
c_{M_i}(t)\leq idSl_{M_i}\cdot\frac{\sum_{j=1}^{i-1}c_{M_j}^{min}-l_{>i}^{max}}{\sum_{j=1}^{i-1}idSl_{M_j}-C}.
\end{equation*}

\rightline{\rule{6pt}{6pt}}

\subsection{Arrival Curve of ST/GB duration}
\label{subsec:AC_ST/GB}
In this section, the arrival curve of ST/GB duration is constructed. It is used to derive the leftover service curve for AVB traffic of Class $M_i$, as AVB traffic cannot be transmitted when credit is frozen. As discussed above, credit will be frozen during ST time slots (windows) depending on the GCL, and whether it is frozen during GB intervals depends on the behavior selection.

\begin{figure}[!t]
	\centering
	\includegraphics[width=0.44\textwidth]{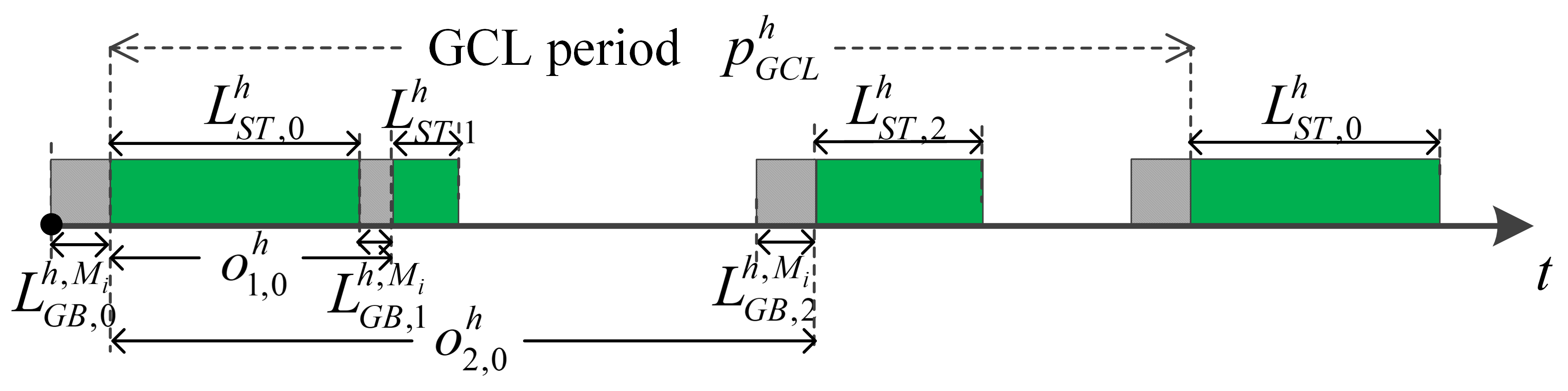}
	\caption{\label{fig:GBttWindow} Guard bands and ST windows}
\end{figure}

ST traffic is scheduled within specific ST windows according to GCLs. The GCL for the output port $h$ is repeated after the GCL period $p^{h}_{\GCL}$, 
see the example in Fig.~\ref{fig:GBttWindow}. Let $N_{ST}^h$ be the number of ST windows in the GCL period $p^{h}_{\GCL}$. It is assumed that the 
$i$th 
ST window in output port $h$ starts at $o^h_i$ and has duration $L_{ST,i}^h$, and the relative offset between the starting time of the $i$th and $j$th 
ST windows 
is $o_{j, i}^h=o^h_j-o^h_i$ 
For the case of credit non-frozen for AVB traffic during guard bands, the arrival curve for the credit frozen part is only related to ST windows, which is given by the following Lemma.
	\begin{lemma}\label{nprAggArrST}
		The arrival curve of the credit frozen part due to ST traffic in an output port $h$ is given by, for all $t\in \mathbb{R}^+$
		\begin{equation}\label{g:alpha_ST}
		\begin{split}
		\alpha_{ST}^{h}(t)&=\mathop{max}_{0\leq i\leq N^h_{ST}-1}\left\{\alpha_{ST,i}^{h}(t)\right\}\\
		\alpha_{ST,i}^{h}(t) &=\sum_{j=i}^{i+N^h_{ST}-1}L_{ST,j}^h C\cdot
		\left\lceil\frac{t-o_{j,i}^h}{p_{\GCL}^h}\right\rceil,
		\end{split}
		\end{equation}
	\end{lemma}
where is $\alpha_{ST,i}^{h}(t)$ is one possible arrival curve by selecting $i$th ($i\in\left[0,N_{ST,i}^h\right]$) ST window as the reference; $L_{ST,j}^h\cdot C$ represents the maximum number of bits the could be transmitted during the ST traffic window of length $L_{ST,j}^h$; each staircase function shows the upper bound of ST transmission in the periodic ST windows of length $L_{ST,j}^h$, and the relative offsets $o_{j,i}^h$ give the relationships between different ST windows within the GCL period. The proof of the lemma is similar to the proof for TT flows in TTEthernet~\cite{Zhao17}.

Moreover, for the case of credit frozen for AVB traffic of Class $M_i$ during GB, the arrival curve for the credit frozen part is not only related to ST windows, but also to GB intervals. A guard band may appear before a ST window, depending on the flow backlog. In the worst-case, the time duration of the guard band $L_{GB,j}^{h,M_i}$ before the $j$th ($j\in[0,N^h_{ST}-1]$) ST window equals to the minimum value of the maximum transmission time ($l_{\leq i}^{max}/C$) of AVB frames of Class $M_i$ competing in output port $h$ and the idle time interval between two consecutive ST windows ($(j-1+N^h_{ST})\%N^h_{ST}$th and $j$th windows). We merge the guard band and ST window together to construct the arrival curve of the credit frozen part, the following Lemma is given,

\begin{lemma}\label{nprAggArrSTGB}
	The arrival curve of the credit frozen part due to ST traffic and guard band in an output port $h$ is given by, for all $t\in \mathbb{R}^+$
	\begin{gather}\label{g:alpha_GB+ST}
	\alpha_{GB+ST}^{h,M_i}(t)=\mathop{max}_{0\leq j\leq N^h_{ST}-1}\left\{\alpha_{GB+ST,j}^{h,M_i}(t)\right\}\\
	\alpha_{GB+ST,j}^{h,M_i}(t) =\sum_{k=j}^{j+N^h_{ST}-1}(L_{ST,k}^h+L_{GB,k}^{h,M_i}) C
	\left\lceil\frac{t-o_{k,j}^h+L_{GB,k}^{h,M_i}-L_{GB,j}^{h,M_i}}{p_{\GCL}^h}\right\rceil,\notag
	\end{gather}
where is $\alpha_{GB+ST,j}^{h,M_i}(t)$ is one possible arrival curve by selecting the $j$th ($j\in\left[0,N_{ST}^h-1\right]$) ST window as the reference.
\end{lemma}

In the following, we are interested to derive $\sigma_{GB}^{M_i}$ and $\rho_{GB}^{M_i}$ defined in Eq.~(\ref{g:deltaTGB}) which is used for deriving credit upper bound of AVB Class $M_i$. They are parameters (burst and long-term rate) related to the upper envelope of accumulated bits in guard band duration $\Delta t_{GB}(s,t)$ in any interval $[s,t]$. Additionally, note that $\sigma_{GB}^{M_i}$ and $\rho_{GB}^{M_i}$ are parameters defined based on $t-s-\Delta t_{ST}(s,t)$ in Eq.~(\ref{g:deltaTGB}). The following Theorem gives the upper envelope of the guard band defined in Eq.~(\ref{g:deltaTGB}) in the form of the staircase function.
\begin{theorem}\label{nprAggArrGBstair}
	The staircase upper bound on the GB duration defined from Eq.~(\ref{g:deltaTGB}) for AVB Class $M_i$ traffic for all $t\in \mathbb{R}^+$ is as follows,
	\begin{equation}\label{g:alpha_GB_a}
	\alpha_{GB}^{h,M_i}(t)= \max_{0\leq j\leq N^h_{ST}-1}\left\{\alpha_{GB,j}^{h, M_i}\left(t-\Delta t_{ST}(t)\right)\right\},
	\end{equation}
	where
	\begin{multline}
	\label{g:alpha_GB_b}
	\alpha_{GB,j}^{h, M_i}\left(t-\Delta t_{ST}(t)\right)=\\
	\sum_{k=j}^{j+N^h_{ST}-1} L_{GB,k}^{h, M_i} C\left\lceil\frac{t-o_{k,j}^h+L_{GB,k}^{h, M_i}+\sum_{p=j}^{k} L_{ST,p}^h}{p_{\GCL}^h-\sum_{p=j}^{j+N^h_{ST}-1} L_{ST,p}^h}\right\rceil.
	\end{multline}
\end{theorem}

\emph{Proof:}
We need to upper bound $\UGB(s,s+t)$ the number of $\GBmax{j}$ in $[s,s+t]$.
Introduce $\forall j: u_j=o_j-\GBmax{j}$.
First, consider some $u_j$ and $t\geq 0$, let $k=\max\set{p\st u_p < u_j+t}$
\begin{align}
\UGB(u_j,u_j+t) 
\label{lem:GBbound:uj:GBmaxkbound}
&\leq \GBmax{j} + \ldots + \GBmax{k-1} + \GBmax{k}
\\
& =\sum_{k=j}^{\infty} 
\GBmax{k} \test{t > u_k-u_j}
\end{align}
Noticing that $\DtST(u_j,u_k)=\sum_{p=j}^{k-1}\Lclose{p}$, Chasle's relation $\DtST(u_j,u_j+t)=\DtST(u_j,u_k)+\DtST(u_k,u_k+t)$ and $\DtST(u_k,u_j+t)\leq \Lclose{k}$ leads to
\begin{align}
\label{eq:DtSTui:bound:infinite}
\UGB(u_j,u_j+d)  \leq 
\sum_{k=j}^\infty
\GBmax{k}
\test{t - \DtST(u_j,u_j+d) > u_k-u_j-\sum_{p=j}^{k} \Lclose{p}}
\end{align}

But any infinite sum on any expression $E(k)$, can be decomposed as a double sum using  $k=k'+n\NGCL$ with $k'=k\textit{ mod }\NGCL$
\begin{align}
\sum_{k=j}^{\infty} E(k) 
= \sum_{k'=j}^{j+\NGCL-1} \sum_{n=0}^\infty  E(k'+n\NGCL)
\end{align}
The system behavior is periodic, so  $\forall k, u_{k+\NGCL}  = u_{k}+\pGCL$, 
$\Lclose{k+\NGCL}=\Lclose{k}$,
$\GBmax{k+\NGCL} = \GBmax{k}$, leading to, $\forall t$
\begin{multline}
\sum_{k=j}^\infty
\test{t > u_k-u_j-\sum_{p=j}^{k}\Lclose{p}}\\
=  \sum_{k'=j}^{j+\NGCL-1} \sum_{n=0}^\infty 
\test{t > u_{k'}+n\pGCL-u_j-\sum_{p=j}^{k'}\Lclose{p}
	- n \LclosePeriod}
\end{multline}
with $\LclosePeriod=\sum_{k=0}^{\NGCL-1}\Lclose{k}$.
Now, remark that $\forall t,J,P\in\nnR$, $P>0$,
\begin{equation}
\label{eq:ceil-as-inf-sum}
\nnfceil{t-J}{P}  = \sum_{n=0}^\infty \test{t>J+nP}.
\end{equation}
Combining all leads to
\begin{multline*}
\sum_{k=j}^\infty \GBmax{k} \test{t > u_k - u_j -\sum_{p=j}^{k} \Lclose{p}}  
\end{multline*}
\begin{multline}
= \sum_{k=j}^{j+N^h_{ST}-1} \GBmax{k} 
\nnfceil{t-(u_k-u_j-\sum_{p=j}^{k} \Lclose{p})}{p^h_{\GCL}-\LclosePeriod}
\isdef F_j(t)
\label{eq:Fjdef}    
\end{multline}

Let $s \geq 0$, it exists $o_j$ such that $s\in[o_{j-1},o_{j})$.
If $s\in [o_{j-1},u_j]$, $\UGB(s,s+t)=\UGB(u_j,s+t)$. Using eq.~\eqref{eq:DtSTui:bound:infinite}, \eqref{eq:Fjdef}, $\UGB(s,s+t)\leq F_j(s+t-u_j-\DtST(u_j,s+t))$. From Chasles's relation and $\DtST(s,u_j)\leq u_j-s$ comes $ s - u_j - \DtST(u_j,s+t) \leq - \DtST(s,s+t)$. And since $F_j$ is non decreasing, it leads to
\begin{align}
\UGB(s,s+t)
\leq F_j(t - \DtST(s,s+t) )
\end{align}
If $s\in [u_j,o_j)$, $\UGB(s,s+t)=\UGB(u_j,s+t)-(s-u_j)$. But for any $x,y,z$, $z \leq y$ $\UGB(x,x+y)-z\leq \UGB(x,x+y-z)$, so $\UGB(s,s+t)\leq \UGB(u_j,u_j+t) \leq F_j(t-\DtST(u_j,u_j+t)).$ Consider now $\DtST$: from $s\in [u_j,o_j]$, 
$\DtST(u_j,u_j+t)=\DtST(s,u_j+t)$,
and from Chalses's relation
$ \DtST(s,s+t)=\DtST(s,u_j+t)+\DtST(u_j+t,s+t)$.
Since $s\in [u_j,o_j]$, $s-u_j \leq o_j-u_j = \GBmax{j}$, then 
$\DtST(u_j+t,s+t) \leq \DtST(u_j+t,o_j+t) 
\leq \GBmax{j}$. So,$ -\DtST(u_j,u_j+t)
\leq - \DtST(s,s+t) + \GBmax{j}$,
and since $F_j$ is non decreasing,
\begin{equation}    
\UGB(s,s+t)
\leq F_j(t+\GBmax{j} -\DtST(s,s+t)).
\end{equation}  
Last, to get rid of the position of $s$ wrt $o_i$ and $u_i$, one gets the maximum
\begin{align}
\UGB(s,s+t) & \leq \max_{j\in\mathbb{N}}\set{F_j(t+\GBmax{j} -\DtST(s,s+t))} \\
= & \max_{0 \leq j \leq N^h_\textit{ST}-1}\set{F_j(t+\GBmax{j} -\DtST(s,s+t))} 
\end{align}{}
Last, we are looking for a bound on $C\cdot\DtST(s,s+t)$ and 
$t+\GBmax{j}-u_k+u_j
= t+\GBmax{j} - (o^h_k-\GBmax{k}) + o^h_j - \GBmax{j}
=t-o^h_{k,j}+\GBmax{k},$ leading to eq.~\eqref{g:alpha_GB_b}.

\rightline{\rule{6pt}{6pt}}

\begin{lemma}\label{nprAggArrGBlinear}
	(Linear bound on staircase function) Let $P\in \mathbb{R}^+$, $\{o_1, ..., o_n\}$, $\{l_1, ..., l_n\} \in \mathbb{R}^+$ be two sets of non-negative values such that $o_1\leq o_2\leq...\leq o_n\leq P$. Let $f(t)= \sum_{j=1}^n l_j\left\lceil\frac{t-o_j}{P}\right\rceil$. Then for all $t\in \mathbb{R}^+$
	\begin{equation}\label{g:alpha_GBlinear}
	f(t) \leq \rho t + \max\set{\sigma_1,...\sigma_n},
	\end{equation}
	with $\sigma_k=\sum_{j=1}^k l_j-\rho o_k$, $\rho=\sum_{j=1}^n l_j/P$. 
\end{lemma}

\begin{figure}[!t]
	\centering
	\includegraphics[width=0.26\textwidth]{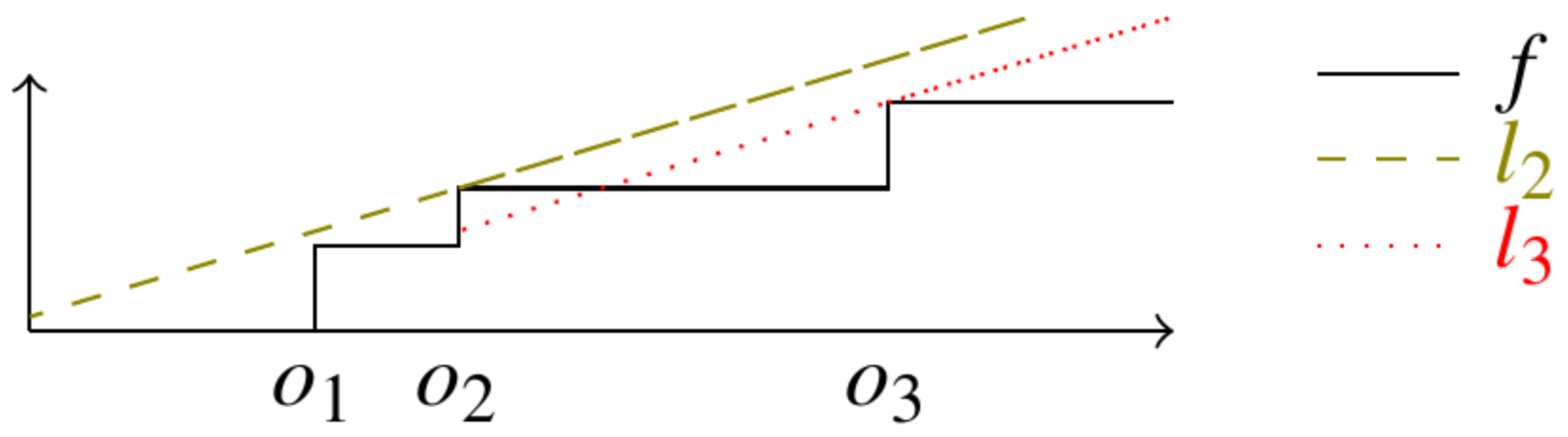}
	\caption{\label{fig:LinBoundStairCase} Linear bound on staircase function}
\end{figure}

\emph{Proof:} For any $k$, let $l_k: t \mapsto \sigma_k + \rho t$. Since $f(t+P)=f(t)+\rho P$, if $\forall u\in[0,P]$, $f(u)\leq l_k(u)$, then $f \leq l_k$. Notice that $l_{k}(o_k)=f(o_k)$, and  $\forall u\in[o_k,o_{k+1}]$ $f(u) \leq \max\set{l_k,l_{k+1}}$ (cf. Fig.~\ref{fig:LinBoundStairCase}). So, $\forall u\in[0,P]$, $f(u)\leq \max_{k\in[1,n]}\set{l_k(u)}$.

\rightline{\rule{6pt}{6pt}}


\subsection{Tighter latency bounds by introducing shaping curves for arrival of AVB traffic}
\label{subsec:Tighter latency & shaping curves}

According to Network Calculus, the upper bound latency of a Class $M_i$ flow $\tau_{M_i[k]}$ in output port $h$ is given by the maximum horizontal deviation between the aggregate arrival curve $\alpha_{M_i}^h(t)$ of intersecting flows of AVB Class $M_i$ and the service curve $\beta_{M_i(I)}^{h}(t)$ for AVB Class $M_i$ in $h$,
\begin{equation}\label{g:17}
D_{M_i[k]}^h=h(\alpha_{M_i}^h(t),\beta_{M_i(I)}^{h}(t)),
\end{equation}
\noindent
where the service curve $\beta_{M_i(I)}^{h}(t)$ is from Theorem~\ref{nprSerAVB}.

The individual arrival curve for the flow $\tau_{M_i[k]}$ in the source ES ($h_0$) can be given by
\begin{equation}
\alpha_{M_i[k]}^{h_0}(t)=l_{M_i[k]}+\frac{l_{M_i[k]}}{p_{M_i[k]}}\cdot t
\end{equation}
\noindent
where $l_{M_i[k]}$ is the burst of the flow $\tau_{M_i[k]}$ in $h_0$, $l_{M_i[k]}/p_{M_i[k]}$ is the long-term rate of $\tau_{M_i[k]}$ sending from $h_0$. The individual arrival curve $\alpha_{{M_i}[k]}^h(t)$ for $\tau_{{M_i[k]}}$ in the intermediate node is calculated from the previous node port $h'$ along the path of $\tau_{M_i[k]}$,
\begin{equation}\label{g:OutArr}
\alpha_{M_i[k]}^{h}(t)=\alpha_{M_i[k]}^{h'}(t)\oslash \delta_{D_{M_i[k]}^{h'}}(t),
\end{equation}
where $D_{M_i[k]}$ is the maximum queuing latency of the flow $\tau_{M_i[k]}$ in output port of $h'$, and $\delta_{D}(t)$ is the burst-delay function~\cite{LeBoudec01} which equals to 0 if $t\leq D$ and $\infty$ otherwise.

Moreover, we provide a solution that leads to tighter aggregate arrival curves for AVB traffic thus reducing pessimism of worst-case delays for AVB flows in intermediate nodes, by taking flows $\tau_{{M_i}[k]}$ of Class $M_i$ from the same previous node port $h'$ as a group. On one hand, such grouped flows are limited by the physical link speed, meaning that they cannot arrive simultaneously. The shaping curve of physical link is given by,
\begin{equation}\label{g:ShaCurPhyLin}
\sigma_{link}(t)=C\cdot t.
\end{equation}
On the other hand, the grouped flows are constrained by CBS as well. The shaping curve of CBS $\sigma_{M_i}^{h'}(t)$ will be discussed in Sect.~\ref{subsec:Shaping Curve of CBS} with the consideration of ST effects.

Thus the aggregate arrival curve $\alpha_{{M_i},h'}^h(t)$ of grouped flows in the input of $h$ from the preceding node port $h'$ is constrained by all three aspects, \emph{i.e.}, sum of individual output arrival curves from $h'$ (Eq.~(\ref{g:OutArr})), the shaping curve of physical link (Eq.~(\ref{g:ShaCurPhyLin})) and the shaping curve of CBS (Eq.~(\ref{g:shaping curve of CBS})). Then it is given by,
\begin{equation}\label{g:GrArr}
\begin{split}
\alpha_{M_i,h'}^h(t)=&\bigg(\sum_{\tau_{{M_i[k]}}\in [h',h]}\alpha_{M_i[k]}^{h}(t)\bigg)\wedge\left(\sigma_{link}(t)+l_{M_i,h'}^{h,max}\right)\\
&\wedge\left(\sigma_{M_i}^{h'}(t)+l_{M_i,h'}^{h,max}\right),
\end{split}
\end{equation}
where $x\wedge y=min\{x,y\}$, $l_{M_i,h'}^{h,max}$ is the maximum frame size of flows of Class $M_i$ from $h'$ to $h$.

Then the aggregate arrival curve $\alpha_{M_i}^h(t)$ for the output port $h$ is the sum of all grouping arrival curves of Class $M_i$. Moreover, it is assumed that there is no limitation for arriving of flows in the source ES ($h_0$). Then they can simultaneously arrive and thus the aggregate arrival curve $\alpha_{M_i}^{h_0}(t)$ for the port $h_0$ is the sum of all individual arrival curves of Class $M_i$ flows.

By disseminating the computation of latency bounds along the routing of $\tau_{M_i[k]}$, its WCD is obtained by the sum of delays from its source ES to its destination ES,
\begin{equation}\label{g:18}
D_{M_i[k]}=\sum_{h\in dr_{M_i[k]}}D_{M_i[k]}^h+(|h|-1)\cdot d_{tech},
\end{equation}
where $|h|-1$ represents the number of SWs the flow passing through and $d_{tech}$ is the constant technical latency in a SW.

\subsection{CBS Shaping Curve for AVB traffic}
\label{subsec:Shaping Curve of CBS}

\begin{lemma}\label{nprStrSerTT}
With the non-preemption mode, the strict service curve for ST traffic in an output port $h$ is given by,
\begin{equation}\label{g:nprStrSerTT}
\beta_{ST}^{h}(t)=
min_{0\leq i\leq N^h_{ST}-1}\bigg\{\sum_{j=i}^{i+N^h_{ST}-1}\beta_{TDMA}^h(t+t_0,L_{ST,j}^h)\bigg\},
\end{equation}
where
\begin{equation*}
\begin{split}
\beta_{TDMA}^h&(t, L)=C\cdot max\bigg\{\bigg\lfloor\frac{t}{p_{\GCL}^h}\bigg\rfloor L, t-\bigg\lceil\frac{t}{p_{\GCL}^h}\bigg\rceil(p_{\GCL}^h-L)\bigg\},
\end{split}
\end{equation*}
and
\begin{equation*}
t_0=p_{\GCL}^h-L_{ST,j}^h-o_{0,i}^h-o_{j,i}^h.
\end{equation*}
\end{lemma}

\emph{Proof}: As discussed in Sect.~\ref{subsec:Service Curve for AVB traffic}, for the given GCL in an output port, there are $N^h_{ST}$ number of ST windows in the GCL period $p_{\GCL}^h$. Then the service for ST traffic can be taken as $N^h_{ST}$ sets of periodic windows of known lengths and relative offsets.

If considering one set of periodic ($p_{\GCL}^h$) ST windows of length $L_{ST}$, its service in the port $h$ is similar to the TDMA communication~\cite{Ernesto06}. In the worst-case, the service cannot be guaranteed for ST traffic during any time interval $0\leq \Delta t<p_{\GCL}^h-L_{ST}^h$, but can be guaranteed of $C\cdot(\Delta t-p_{\GCL}^h-L_{ST}^h)$ in any time interval $p_{\GCL}^h-L_{ST}^h\leq \Delta t<p_{\GCL}^h$. Then the service curve for periodic ST windows during $\Delta t$ can be given by,
\begin{equation}\label{g:TDMA}
\begin{split}
\beta&_{TDMA}^h(\Delta t, L_{ST}^h)\\
&=C\cdot max\bigg\{\bigg\lfloor\frac{\Delta t}{p_{\GCL}^h}\bigg\rfloor L_{ST}^h, \Delta t-\bigg\lceil\frac{\Delta t}{p_{\GCL}^h}\bigg\rceil(p_{\GCL}^h-L_{ST}^h)\bigg\}.
\end{split}
\end{equation}

Then the service for all $N^h_{ST}$ sets of periodic ST windows is derived from (\ref{g:TDMA}) as well as the relative offsets between two ST windows from different sets. Taking the $i$th ($i\in[0,N^h_{ST}-1]$) ST window as the reference, we have that the first served frame is from the $i$th ST window during the backlogged period $\Delta t$. Assume that $o_{0,i}^h$ is the maximum idle time interval before the opening time of the $i$th window, then the service guaranteed for the $i$th set of ST windows is the curve in (\ref{g:TDMA}) shifted to the left with the positive value $p_{\GCL}^h-L_{ST,i}^h-o_{0,i}^h$,
\begin{equation*}
\beta_{ST,i,i}^h(\Delta t)=\beta_{TDMA}^h(\Delta t+p_{\GCL}^h-L_{ST,i}^h-o_{0,i}^h,L_{ST,i}^h).
\end{equation*}
Moreover, for the other set of the $j$th ($j\in [i+1,i+N^h_{ST}-1]$) traffic windows, with the known of the relative offset $o_{j,i}^h$ by considering the $i$th ST window as the benchmark, the service guaranteed for ST traffic in the $j$th set of windows can be given by shifting the curve in (\ref{g:TDMA}) to the left with the positive value $p_{\GCL}^h-L_{ST,j}^h-o_{0,i}^h-o_{j,i}^h$,
\begin{equation*}
\beta_{ST,j,i}^h(\Delta t)=\beta_{TDMA}^h(\Delta t+p_{\GCL}^h-L_{ST,j}^h-o_{0,i}^h-o_{j,i}^h,L_{ST,j}^h).
\end{equation*}
Note that $\beta_{ST,j,i}^h(\Delta t)$ equals to $\beta_{ST,i,i}^h(\Delta t)$ if $j=i$.

Thus if the $i$th ST window is as benchmark, the strict service for ST traffic is as follows after considering all $N^h_{ST}$ sets of periodic ST windows,
\begin{equation*}
\beta_{ST,i}^h(\Delta t)=\sum_{j=i}^{i+N^h_{ST}-1}\beta_{ST,j,i}^h(\Delta t).
\end{equation*}
Then the strict service curve for ST traffic in an output port $h$ is the lower envelope of $\beta_{ST,i}^{h}(\Delta t)$ by considering each ST windows in the GCL period as benchmark,
\begin{equation*}
\beta_{ST}^h(\Delta t)=min_{0\leq i\leq N^h_{ST}-1}\{\beta_{ST,i}^{h}(\Delta t)\}.
\end{equation*}

\rightline{\rule{6pt}{6pt}}

\begin{theorem}\label{Shaping Curve of CBS}
The CBS shaping curve of Class $M_i$ for the non-preemption mode is given by,
\begin{equation}\label{g:shaping curve of CBS}
\sigma_{M_i}^{h}(t)=\bigg[t-\frac{\beta_{ST}^h(t)}{C}\bigg]_\uparrow^+\cdot idSl_{M_i}+c_{M_i}^{max}-c_{M_i}^{min},
\end{equation}
\end{theorem}
where $\beta_{ST}^h(t)$ is given by the Lemma~\ref{nprStrSerTT}, and $c_{M_i}^{max}$ and $c_{M_i}^{min}$ are respectively the upper and lower bounds of credit of $M_i$ given by Theorems~\ref{credit_upper_bound} and \ref{credit_lower_bound}.

\emph{Proof}: For an arbitrary period of time $\Delta t$, the variation of credit during $\Delta t$ satisfies
\begin{equation}\label{g:credit variation}
\begin{split}
\Delta c_{M_i}&=c_{M_i}(t+\Delta t)-c_{M_i}(t)=\Delta t^+\cdot idSl_{M_i}+\Delta t^-\cdot sdSl_{M_i} \\
&=(\Delta t-\Delta t^0)\cdot idSl_{M_i}-\Delta t^-\cdot(idSl_{M_i}-sdSl_{M_i}).
\end{split}
\end{equation}

\noindent
In the best-case, the frozen duration $\Delta t^0=\Delta t^{ST}$ is only related to ST windows, and has nothing to do with guard bands, for the non-preemption integration mode. Considering the strict service curve of ST traffic expressed in the Lemma \ref{nprStrSerTT}, $\Delta t^{ST}$ is limited by,
\begin{equation}\label{g:24}
R_{ST}^{h*}(t+\Delta t)-R_{ST}^{h*}(t)=\Delta t^{ST}\cdot C\geq \beta_{ST}^h(\Delta t).
\end{equation}
Moreover, $\Delta t^-$ can be expressed by,
\begin{equation}\label{g:25}
R_{M_i}^{h*}(t+\Delta t)-R_{M_i}^{h*}(t)=\Delta t^-\cdot C,
\end{equation}
and $\Delta c_{M_i}$ satisfies the relationship as follows,
\begin{equation}\label{g:26}
\Delta c_{M_i}\geq c_{M_i}^{min}-c_{M_i}^{max}.
\end{equation}
Due to non-decreasing function $R_{M_i}^{h*}(t)$ and using the expressions (\ref{g:24}), (\ref{g:25}), (\ref{g:26}) and $sdSl_{M_i}=idSl_{M_i}-C$, (\ref{g:credit variation}) is modified to obtain,
\begin{equation*}
\begin{split}
R_{M_i}^{h*}(t+\Delta t)-R_{M_i}^{h*}(t)\leq \bigg[\Delta t-\frac{\beta_{ST}^h(\Delta t)}{C}\bigg]_\uparrow^+\cdot idSl_{M_i}+c_{M_i}^{max}-c_{M_i}^{min}.
\end{split}
\end{equation*}

\rightline{\rule{6pt}{6pt}}

\section{Experimental Results}
\label{sec:Experimental Results}

We have evaluated our proposed improving Network Calculus-based WCD analysis for multiple AVB classes with consideration of both frozen and non-frozen credit during guard band (GB) in TSN (called impNC/TSN) as follows. impNC/TSN is implemented in C++ using the Java kernel of the RTC toolbox~\cite{Wanderler06}, running on a computer with Intel Core i7-8550U CPU at 1.80 GHz and 8 GB of RAM.


\subsection{Synthetic Test Cases}
In this section, our evaluation focuses on a topology of 6 ESes and 2 SWs, connected via physical links with rates of 100Mb/s. The test case (TC1) has 12 ST flows and 10 AVB flows with two classes ($M_1$ and $M_2$). Only two classes of AVB traffic are considered in TC1, as we are interested in comparing our improved method (impNC/TSN) with the other existing methods~\cite{Maxim17,ZhaoRTAS17} from related work, which only support the analysis for two classes of AVB traffic. The average load of links is about 14.4\% and the maximum link load is about 33.7\%. The idle slopes of Class $M_1$ and Class $M_2$ are respectively set to 40\% and 20\% \footnote{More detailed information for TC1 and TC2 can be downloaded from \url{https://zenodo.org/record/3785915#.XrDsH2j7RPZ}}.

Firstly, we are interested to show the improvement of AVB WCDs in TSN network obtained with our proposed method by considering physical link and CBS shaping curves in comparison to existing work~\cite{Maxim17,ZhaoRTAS17}. The earliest method proposed for AVB analysis in TSN network is based on Eligible Interval Analysis~\cite{Maxim17}, which we denote with EIA17. The NC-based method from Zhao~\cite{ZhaoRTAS17} without shaping constraints is denoted with NC/TSN18. Additionally, we give simulation results (denoted as Sim) in order to show the correctness of the proposed method to some extent. Note that here we use the same assumption of the credit variation as related work, i.e., credit frozen during guard band.


\begin{table}
\scriptsize
\caption{Compared results by different methods}
\label{tab:2}
\centering
\begin{tabular}{|c|c|c|c|c|}
\hline
Flow & \tabincell{c}{EIA17 ($\mu$s)} & \tabincell{c}{NC/TSN18 ($\mu$s)} & \tabincell{c}{impNC/TSN ($\mu$s)} & \tabincell{c}{Sim ($\mu$s)}\\
\hline\hline
AVB1  & 3798 & 3449 & 2456  & 1448 \\
\hline
AVB2  & 3778 & 3273 & 2379  & 1559 \\
\hline
AVB3  & 4771 & 3663 & 2466  & 1529 \\
\hline
AVB4  & 5230 & 5635 & 4442  & 2699 \\
\hline
AVB5  & 4971 & 4612 & 3865  & 1721 \\
\hline
AVB6  & 5151 & 5287 & 4421  & 2615 \\
\hline
AVB7  & 4969 & 3744 & 2331  & 1306 \\
\hline
AVB8  & 4642 & 4236 & 3490  & 1996 \\
\hline
AVB9  & 4757 & 5635 & 4442  & 2494 \\
\hline
AVB10 & 5005 & 5287 & 4421 & 2436 \\
\hline
\end{tabular}
\end{table}

The results for each AVB flow of TC1 are presented in Table~\ref{tab:2}, calculated from different timing analysis methods. As we can see, compared with our impNC/TSN method, EIA17 suffers the larger WCDs, with 43.6\% larger on average. This is expected, since EIA17 does not consider the relative locations of ST windows. The pessimism of EIA17 will be even larger with the increasing interval among ST windows. Compared with NC/TSN18, our method by introducing physical link shaping and CBS shaping curves is able to significantly reduce the WCDs compared to \cite{ZhaoRTAS17}. As we can see from the third and forth columns in Table~\ref{tab:2}, impNC/TSN is able to reduce on average the WCDs obtained with~\cite{ZhaoRTAS17} by 23.5\% and as much as 37.7\% in some cases. Additionally, we give simulation results in the last column. The simulation results denoted with Sim are useful for validating our approach, as our obtained WCDs should all be larger than the maximum delays obtained by simulation, which is the case in our experiments. However, as rare events can be missed, a simulation approach will not be able to determine the WCDs, hence it is not useful for safety critical applications that require safety guarantees.

\begin{figure}[t]
	\centering
	\includegraphics[width=0.41\textwidth]{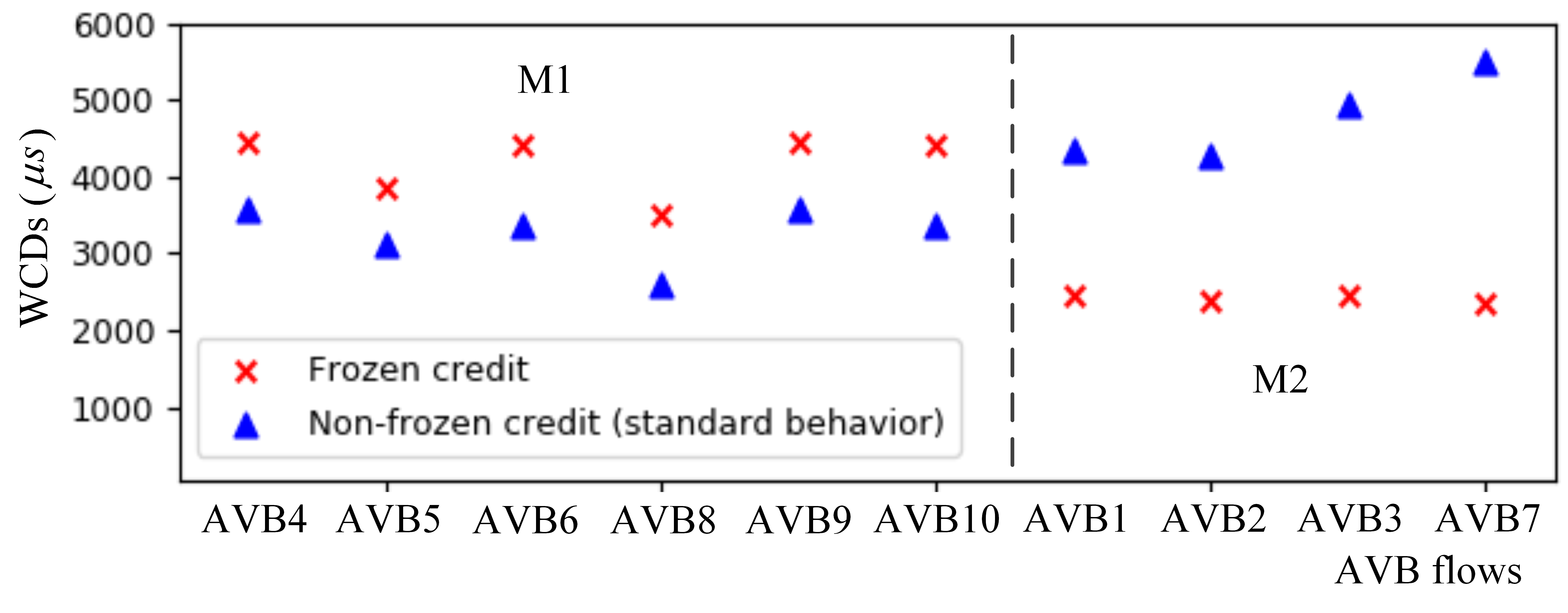}
	\caption{\label{fig:frozen-nonfrozen}Comparison of WCDs with frozen and non-frozen credit during GB}
\end{figure}

For the second experiment, to show the impact of frozen and non-frozen credit during GB on the WCD of AVB traffic, we use the same TC1, but just assume different credit variations during guard band. We show the results in Fig.~\ref{fig:frozen-nonfrozen}. The values on the x-axis give the AVB flows, from AVB1 to AVB10, and on the y-axis show the WCDs in microseconds. 
As it can seen in Fig.~\ref{fig:frozen-nonfrozen}, WCDs of Class $M_1$ under frozen credit assumption are larger than the Class $M_1$ results under the non-frozen credit assumption, while WCDs of Class $M_2$ are just the opposite. This confirms our intuition, since if considering non-frozen credit, both credits for Class $M_1$ and Class $M_2$ will be increased during GB. Moreover, as Class $M_2$ has lower priority than Class $M_1$, the increased possibility of $M_1$ obtaining service means that the possibility of $M_2$ getting service is reduced.

\subsection{Evaluation on a Larger Realistic Test Case}

In this section, we use a larger real-world test case (TC2), adapted from the Orion Crew Exploration Vehicle (CEV)~\cite{TamasCSelicean15} by using the same topology and ST flows, and considering rate-constrained (RC) flows as AVB flows. We investigate the scalability of our method, show the extension of the analysis method on multiple AVB classes and investigate the influence of varied idle slopes on multiple classes of AVB traffic. Without loss of generality, the credit is assumed to be frozen during GB in this subsection. CEV has a topology consisting of 31 ESes, 15 SWs, 188 dataflow routes, 
connected by dataflow links transmitting at 1 Gbps, and running 100 ST flows, 25 AVB flows of Class $M_1$, 25 AVB flows of Class $M_2$, 24 AVB flows of Class $M_3$ and 13 AVB flows of Class $M_4$ (including multicast flows).

\begin{figure}[t]
	\centering
	\includegraphics[width=0.5\textwidth]{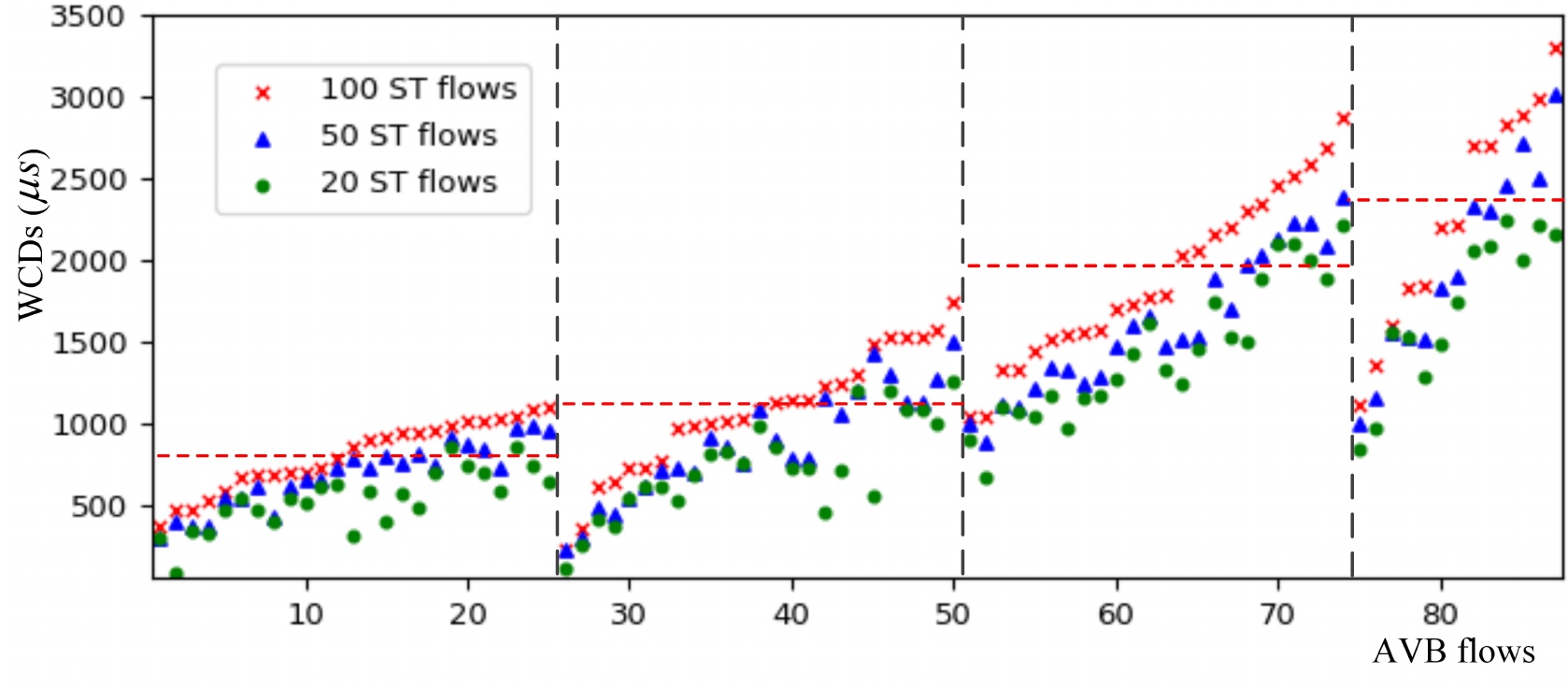}
	\caption{\label{fig:5-50TT}Compared WCDs with different number of ST flows}
\end{figure}

In such experiment, we would like to evaluate the scalability of our analysis impNC/TSN with the number of ST flows. We reduce the total number of ST flows from 100 to 50 and 20 respectively. The idle slope for Classes $M_i$ ($i=1, 2, 3, 4$) are respectively 40\%, 20\%, 10\% and 5\%. The obtained results are visually shown in Fig.~\ref{fig:5-50TT}
As shown in Fig.~\ref{fig:5-50TT}, the obtained results are grouped by the AVB Classes with vertical dotted lines and respectively sorted in increasing order by results from the case with 100 ST flows. 
As expected and can be seen from Fig.~\ref{fig:5-50TT}, the WCDs of the AVB flows grow with the increasing number of ST flows. In addition, the average WCD (horizontal dotted lines) of Class $M_i$ is smaller than $M_{i+1}$, since the AVB class of higher priority has larger bandwidth guarantee, i.e., $idSl_{M_i}>idSl_{M_{i+1}}$.

\section{Conclusion}
\label{sec:Conclusion and Future Work}
The TSN IEEE task group is defining new extensions of Ethernet, devoted to real-time and safety-critical application areas. These types of applications require a method to bound the worst-case delays of a given configuration. This paper has presented an improved Network Calculus-based approach by considering the limitations from physical link rate and the output of CBS, which are modeled as shaping curves to compute tighter bounds. Moreover, the timing analysis method has be extended to arbitrary number of AVB traffic classes, considering both frozen and non-frozen of credit during guard band, in the case of a TSN/GCL+CBS architecture.

Our analysis is, to the best of our knowledge, the first one to handle with both credit frozen and non-frozen during guard bands and with any number of AVB classes under the influence of ST traffic for the whole TSN network. We have evaluated the proposed approach on both synthetic and realistic test cases. The comparison results to the existing approaches show that the Network Calculus approach is a viable approach for the analysis of TSN. Our approach provides safe upper bounds on WCDs, reduces the pessimism of the analysis (tighter WCD bounds), and is scalable to handle large problem sizes.


%





\ifCLASSOPTIONcaptionsoff
  \newpage
\fi


\begin{thebibliography}{1}

\bibitem{IEEE802.3}
IEEE, ``802.3 Standard for Ethernet,'' 2015.

\bibitem{Decotignie05}
J. D.~Decotignie, ``Ethernet-based real-time and industrial communications,'' \emph{Proceedings of the IEEE}, vol. 93, no. 6, pp. 1102-1117, 2005.

\bibitem{ARINC03}
ARINC 664, Aircraft Data Network, Part 7: Deterministic networks, 2003.

\bibitem{SAE11}
SAE AS6802: Time-Triggered Ethernet, Technical report, 2011.

\bibitem{Jansen04}
D.~Jansen, and B.~Holger, ``Real-time Ethernet: the EtherCAT solution,'' \emph{Computing and Control Engineering}, vol. 15, no. 1, pp. 16-21, 2004.


\bibitem{IEEETSNTG}
IEEE, ``Time-Sensitive Networking Task Group,'' \url{http://www. ieee802.org/1/pages/tsn.html}, 2016.

\bibitem{IEEE802.1Q}
IEEE, ``802.1Q---IEEE Standard for Local and Metropolitan Area Networks---Bridges and Bridged Networks,'' \url{https://standards.ieee.org/standard/802_1Q-2018.html}, 2018.

\bibitem{IEEEBA}
IEEE, ``802.1BA---Audio Video Bridging (AVB) Systems,'' \url{http://www.ieee802.org/1/pages/802.1ba.html}, 2011.

\bibitem{IEEEQbv}
IEEE, ``802.1Qbv---Enhancements for Scheduled Traffic,'' \url{http://www.ieee802.org/1/pages/802.1bv.html}, 2015.

\bibitem{IEEEASrev}
IEEE, ``802.1ASrev---Timing and Synchronization for Time-Sensitive Applications,'' \url{http://www.ieee802.org/1/pages/802.1AS-rev.html}, 2017.

\bibitem{Craciunas16}
S. S.~Craciunas, R.~Serna~Oliver, M.~Chmelik, and W.~Steiner, ``Scheduling Real-Time Communication in IEEE 802.1Qbv Time Sensitive Networks,'' in \emph{Proc. of the 24th Int. Conf. on Real-Time Networks and Systems}, 2016.


\bibitem{LuxiAccess18}
L.~Zhao, P.~Pop, and S. S.~Craciunas, ``Worst-case latency analysis for IEEE 802.1 Qbv time sensitive networks using network calculus,'' \emph{IEEE Access}, vol. 6, pp. 41803-41815, 2018.


\bibitem{ZhangAccess19}
P.~Zhang, Y.~Liu, J.~Shi, Y.~Huang, and Y.~Zhao, ``A Feasibility Analysis Framework of Time-Sensitive Networking Using Real-Time Calculus,'' \emph{IEEE Access}, vol. 7, pp. 90069-90081, 2019.


\bibitem{Azua14}
J. A. R.~Azua, and M.~Boyer, ``Complete modelling of AVB in network calculus framework,'' in \emph{Proc. of the 22nd Int. Conf. on Real-Time Networks and Systems}, 2014.

\bibitem{Philip14}
A.~Philip, D.~Thiele, R.~Ernst, and J.~Diemer, ``Exploiting shaper context to improve performance bounds of ethernet avb networks,'' in \emph{Proc. of the 51st Annual Design Automation Conference}, 2014.



\bibitem{NC-AVB-3classes}
L.~Zhao, F.~He, and E.~Li, ``Improving worst-case delay analysis for traffic of additional stream reservation class in Ethernet-AVB Network,'' \emph{Sensors}, vol 18, no. 11, 2018.

\bibitem{FA-EI-AVB-Mclasses}
J. Y.~Cao, P. J. L.~Cuijpers, R. J.~Bril, and J. J.~Lukkien, ``Independent WCRT analysis for individual priority classes in Ethernet AVB. \emph{Real-Time Systems}, vol. 54, no. 4, pp. 861-911, 2018.



\bibitem{Zhao17}
L.~Zhao, P.~Pop, Q.~Li, J. Y.~Chen, and H. G.~Xiong, ``Timing analysis of rate-constrained traffic in TTEthernet using network calculus,'' \emph{Real-Time Systems}, vol. 53, no. 2, pp. 254-287, 2017.

\bibitem{Thiele15}
D.~Thiele, R.~Ernst, J.~Diemer ``Formal Worst-Case Timing Analysis of Ethernet TSN's Time-Aware and Peristaltic Shapers,'' in \emph{Proc. of the IEEE Vehicular Networking Conference}, 2015.

\bibitem{Mohammadpour18}
E.~Mohammadpour, E.~Stai, M.~Mohiuddin, and J.-Y.~Le Boudec, ``End-to-end Latency and Backlog Bounds in Time-Sensitive Networking with Credit Based Shapers and Asynchronous Traffic Shaping,'' in \emph{Proc. of the 30th International Teletraffic Congress}, 2018.

\bibitem{Zhang19}
J.~Zhang, L.~Chen, T.~Wang, and X.~Wang, ``Analysis of TSN for industrial automation based on network calculus,'' in \emph{Proc. of 34th IEEE Int. Conf. on Emerging Technologies and Factory Automation}, 2019.

\bibitem{Laursen16}
S. M.~Laursen, P.~Pop, and W.~Steiner, ``Routing Optimization of AVB Streams in TSN Networks,'' \emph{ACM Sigbed Review}, vol. 13, no. 4, pp. 43-48, 2016.

\bibitem{Maxim17}
D.~Maxim, and Y. Q.~Song, ``Delay Analysis of AVB traffic in Time-Sensitive Networks (TSN),'' in \emph{Proc. of the 25th Int. Conf. on Real-Time Networks and Systems}, 2017.

\bibitem{ZhaoRTAS17}
L. X.~Zhao, P.~Pop, Z.~Zheng, and Q.~Li, ``Timing Analysis of AVB Traffic in TSN Networks using Network Calculus,'' in \emph{Proc. of IEEE Real Time and Embedded Technology and Applications Symposium}, 2018.

\bibitem{Mohammadpour19}
E.~Mohammadpour, E.~Stai, and J.~Y.~Le~Boudec, ``Improved Credit Bounds for the Credit-Based Shaper in Time-Sensitive Networking,'' \emph{IEEE Networking Letters}, vol. 1, no. 3, 2019.


\bibitem{Daigmorte19}
H.~Daigmorte, and M.~Boyer, ``Impact on credit freeze before gate closing in CBS and GCL integration into TSN,'' in \emph{Proc. of International Conference on Real-Time Networks and Systems}, 2019.

\bibitem{LeBoudec01}
J. Y.~Le Boudec, and P.~Thiran, ``Network calculus: A theory of deterministic queuing systems for the internet,'' \emph{Springer-Verlag Lecture Notes on Computer Science}, 5th ed., New York, 2001.






\bibitem{Ernesto06}
W.~Ernesto, L.~Thiele, ``Optimal TDMA time slot and cycle length allocation for hard real-time systems,'' \emph{Proc. of the 2006 Asia and South Pacific Design Automation Conference}, 2006.

\bibitem{Wanderler06}
E.~Wanderler, and L.~Thiele, ``Real-Time Calculus (RTC) Toolbox,'' http://www.mpa.ethz.ch/Rtctoolbox, 2006.



\bibitem{TamasCSelicean15}
D.~Tamas-Selicean, P.~Pop, and W.~Steiner, ``Design optimization of TTEthernet-based distributed real-time systems,'' \emph{Real-Time Systems}, vol. 51, no. 1, pp.1-35, 2015.

\end{thebibliography}
\end{document}